\documentclass[11pt]{article}
\usepackage{moriond,epsfig}

\def\lsim{\mathrel{\raise.3ex\hbox{$<$\kern-.75em\lower1ex\hbox{$\sim$}}}}
\def\gsim{\mathrel{\raise.3ex\hbox{$>$\kern-.75em\lower1ex\hbox{$\sim$}}}}

\def\be{\begin{equation}}
\def\ee{\end{equation}}
\def\bea{\begin{eqnarray}}
\def\eea{\end{eqnarray}}

\begin{document}
\vspace*{4cm}
\title{
DETERMINATION OF HIGGS-BOSON COUPLINGS AT THE LHC%
\footnote{Talk given by G.~Weiglein}
}

\author{M.~D\"UHRSSEN$^{1}$%
, S.~HEINEMEYER$^{2}$%
, H.~LOGAN$^{3}$%
, D.~RAINWATER$^{4}$,\\[.3em]
G.~WEIGLEIN$^{5}$ and D.~ZEPPENFELD$^{6}$
 }

\address{
$^1$Physikalisches Institut, Universit\"at Freiburg, D--79104 Freiburg, 
Germany\\
$^2$CERN TH Division, Dept.\ of Physics,
CH-1211 Geneva 23, Switzerland\\
$^3$Dept.\ of Physics, University of Wisconsin, Madison, Wisconsin 53706
USA\\
$^4$DESY Theory, Notkestr.\ 85, D--22603 Hamburg, Germany\\
$^5$Institute for Particle Physics Phenomenology, University of
Durham, Durham DH1~3LE, UK\\
$^6$Institut f\"ur Theoretische Physik, Universit\"at Karlsruhe,
D--76128 Karlsruhe, Germany
}

\maketitle\abstracts{
  We investigate the determination of Higgs boson couplings to gauge 
  bosons and fermions at the LHC from data on Higgs boson production and
  decay. We demonstrate that very mild theoretical assumptions, which
  are valid in general multi-Higgs doublet models, are sufficient to
  allow the extraction of absolute values of the couplings rather than
  just ratios of the couplings. For Higgs masses below 200~GeV we
  find accuracies of $10-40\%$ for the Higgs couplings and the total
  Higgs boson width after several years of LHC running. The sensitivity
  of the Higgs coupling measurements to deviations from the Standard Model
  predictions is studied for an MSSM scenario.
}

\section{Introduction}

If the Higgs mechanism is realized in nature, it is very likely that at
least one Higgs boson will be dicovered at the LHC. Within the Standard
Model (SM), the Higgs boson can be observed in a variety of channels, in
particular if
its mass lies in the intermediate mass region, $114<m_H\lsim 250$~GeV,
as suggested by direct searches~\cite{Barate:2003sz} and electroweak
precision data~\cite{Grunewald:2003ij}. The situation is similar for
Higgs bosons in this mass range in many extensions of the SM. Once a 
Higgs-like state is discovered, a precise measurement of its couplings
will be mandatory in order to experimentally verify (or falsify) the
Higgs mechanism.

The various Higgs couplings determine Higgs production cross sections
and decay branching fractions. By measuring the rates of multiple
channels, various combinations of couplings can be determined. A
principal problem at the LHC is that there is no technique analogous
to the measurement of the missing mass spectrum at a linear
collider~\cite{Garcia-Abia:1999kv} which would directly determine the
total Higgs production cross section. In addition, some Higgs decay
modes cannot be observed at the LHC.  For example, $H\to gg$ or decays
into light quarks will remain hidden below overwhelming QCD dijet
backgrounds. The $H \to b \bar b$ decay, which has by far the
dominant branching ratio for a light SM-like Higgs, will be detectable
but suffers from large experimental uncertainties. As a consequence of
the strong correlations in the measurements of different Higgs
couplings, only ratios of couplings (or partial widths) can be determined 
if no additional theoretical are made, see e.g.\ the analysis of
Ref.~\cite{ATL-PHYS-2003-030}. 

It is therefore interesting to investigate whether absolute
determinations of couplings become possible if suitable theoretical
assumptions are employed. In
Refs.~\cite{Djouadi:2000gu,Zeppenfeld:2000td} (see also
Ref.~\cite{Belyaev:2002ua}) such a strategy has been outlined, assuming
the absence of unexpected decay
channels and a SM ratio of the $H\to b\bar{b}$ and $H\to\tau\tau$
partial widths. These assumptions are valid, however, in only a
restricted class of models. They can be violated, for instance, in the
Minimal Supersymmetric Standard Model (MSSM).

In the present analysis~\cite{Hcoupl,Assamagan:2004mu} we make only a very mild
theoretical assumption, which is valid in general multi-Higgs doublet 
models (with or without extra Higgs singlets; this class of models
contains in particular the MSSM). In this class of models the strength
of the Higgs--gauge-boson couplings does not exceed the SM value. We
will demonstrate that the existence of such an upper bound on the 
Higgs--gauge-boson couplings is already sufficient to allow the
extraction of absolute couplings rather than coupling ratios.

We consider
the expected accuracies at various stages of the LHC program: after
30~fb$^{-1}$ of low luminosity ($10^{33}\,{\rm cm}^{-2}{\rm
  sec}^{-1}$) running, 300~fb$^{-1}$ at high luminosity
($10^{34}\,{\rm cm}^{-2}{\rm sec}^{-1}$), and a mixed scenario where
the weak boson fusion channels are assumed to suffer substantially
from pile-up problems under high luminosity running conditions (making
forward jet tagging and central jet veto fairly inefficient).

In order to investigate the sensitivity of the coupling measurements at
the LHC to deviations from the SM predictions we consider as a specific
example the no-mixing benchmark scenario of the MSSM as defined in
Ref.~\cite{Carena:2002qg}.
Other MSSM benchmark scenarios have been analyzed in 
Refs.~\cite{Hcoupl,Assamagan:2004mu}. 


\section{Strategy}

\label{sec:strategy}

In order to determine the properties of a physical state such as a
Higgs boson, one needs at least as many separate measurements as
properties to be measured, although two or more measurements can be
made from the same channel if different information is used, e.g.,
total rate and an angular distribution.  Fortunately, the LHC will
provide us with many different Higgs observation channels.  In the SM
there are four relevant production modes: gluon fusion (GF;
loop-mediated, dominated by the top quark), which dominates
inclusive production; weak boson fusion (WBF), which has an
additional pair of hard and far-forward/backward jets in the final
state; top-quark associated production ($t\bar{t}H$); and weak boson
associated production ($WH,ZH$), where the weak boson is identified by
its leptonic decay.

Although a Higgs boson is expected to couple to all SM particles, not 
all these decays would be observable.  Very rare decays (e.g., to
electrons) would have no observable rate, and other modes are
unidentifiable QCD final states in a hadron collider environment
(gluons or quarks lighter than bottom).  In general, however, the LHC
will be able to observe Higgs decays to photons, weak bosons, tau
leptons and $b$ quarks, in the range of Higgs masses where the
branching ratio (BR) in question is not too small.

For a Higgs in the intermediate mass range, the total width, $\Gamma$,
is expected to be small enough to use the narrow-width approximation
in extracting couplings.  The rate of any channel (with the $H$
decaying to final state particles $xx$) is, to good approximation,
given by
\begin{equation}
\sigma(H) \times {\rm BR} (H\to xx) = 
{\sigma(H)^{\rm SM}\over \Gamma_p^{\rm SM}}
\cdot {\Gamma_p\Gamma_x \over \Gamma}\;,
\end{equation}
where $\Gamma_p$ is the Higgs partial width involving the production
couplings, and where the Higgs branching ratio for the decay is written
as ${\rm BR}(H\to xx)=\Gamma_x/\Gamma$. Even with cuts, the observed
rate directly determines the product $\Gamma_p\Gamma_x/\Gamma$
(normalized to the calculable SM value of this product).  The LHC will
have access to (or provide upper limits on) combinations of
$\Gamma_g,\Gamma_W,\Gamma_Z, \Gamma_\gamma,\Gamma_\tau,\Gamma_b$ and
the square of the top Yukawa coupling, $Y_t$.~\footnote{We do not
  write this as a partial width, $\Gamma_t$, because, for a light
  Higgs, the decay $H\to t\bar{t}$ is kinematically forbidden.}

We use the following channels in our
analysis~\cite{Hcoupl,Assamagan:2004mu}:
GF $gg \rightarrow H \rightarrow Z Z$,
WBF $qq \, H \rightarrow qq\, Z Z$,
GF $gg \rightarrow H \rightarrow W W$,
WBF $qq \, H \rightarrow qq\, W W$,
$W \, H \rightarrow W\, W W$ (2$l$ and 3$l$ final state),
$t\bar{t} \, H (H \rightarrow W W, t \rightarrow W b)$ (2$l$ and 3$l$
final state),
inclusive Higgs boson production: $H \rightarrow \gamma \gamma$,
WBF $qq \, H \rightarrow qq\, \gamma \gamma$,
$t\bar{t} \, H (H \rightarrow \gamma \gamma)$,
$W \, H (H \rightarrow \gamma \gamma)$,
$Z \, H (H \rightarrow \gamma \gamma)$,
WBF $qq \, H \rightarrow qq\, \tau \tau$,
$t\bar{t} \, H (H \rightarrow b\bar{b})$.

The production and decay channels listed above refer to a single
Higgs resonance, with decay signatures which also exist in the SM. The
Higgs sector may be much richer, of course.  For instance, the MSSM 
with its two Higgs doublets predicts the existence of three neutral 
and one pair of
charged Higgs boson, and the LHC may be able to directly observe
several of these resonances.  Within SUSY models, additional decays,
e.g., into very light super-partners, may be kinematically allowed.
The additional observation of super-partners or of heavier Higgs
bosons will strongly focus the theoretical framework and restrict the
parameter space of a Higgs couplings analysis.
For our present analysis we ignore the
information which would be supplied by the observation of additional
new particles. Instead we ask the question of how well
LHC measurements of the above decay modes of a single Higgs resonance
can determine the various Higgs boson couplings or partial widths.


While from the channels listed above ratios of couplings (or partial
widths) can be extracted in a fairly model-independent way, see e.g.\
Ref.~\cite{ATL-PHYS-2003-030}, further theoretical assumptions are
necessary in order to determine absolute values of the Higgs couplings
to fermions and bosons and of the total Higgs boson width. The only
assumption that we will make in the following is that the strength of
the Higgs--gauge-boson couplings does not exceed the SM value,
\begin{equation}
\Gamma_V\leq\Gamma_V^{\rm SM}, \quad V=W,Z~.
\label{eq:constraint}
\end{equation}
This assumption is justified in any model with an arbitrary number of
Higgs doublets
(with or without additional Higgs singlets), i.e., it is true for the
MSSM in particular.

While eq.~(\ref{eq:constraint}) constitutes an upper bound on the Higgs
coupling to weak bosons, the mere observation of Higgs production
puts a lower bound on the production couplings and, thereby, on the
total Higgs width. The constraint $\Gamma_V\leq\Gamma_V^{\rm SM}$,
combined with a measurement of $\Gamma_V^2/\Gamma$ from observation of
$H\to VV$ in WBF, then puts an upper bound on the Higgs total width,
$\Gamma$. Thus, an absolute determination of the Higgs total width is
possible in this way. Using this result, an absolute determination also
becomes possible for Higgs couplings to gauge bosons and
fermions.


We obtain the expected LHC accuracies from a fit based on 
experimental information for the channels listed above. For details of
the fitting procedure, see Refs.~\cite{ATL-PHYS-2003-030,Hcoupl}.
The statistical errors for the results presented in 
Sec.~\ref{sec:results} are obtained for the case that the channels
listed above are observed with SM rates. In the fit we allow for
undetected Higgs decays (giving rise to additional partial widths) 
and additional
contributions to the loop-induced Higgs couplings to photon pairs or
gluon pairs due to non-SM particles running in the loops. 
The estimated systematic errors~\cite{ATL-PHYS-2003-030,Hcoupl} include a
$5\%$ luminosity error, uncertainties on the
reconstruction/identification of leptons ($2\%$), photons ($2\%$),
b-quarks ($3\%$), $\tau$-jets ($3\%$) and forward tagging jets and
veto jets ($5\%$), error propagation for background determination from
side-band analyses (assuming an error from $0.1\%$ for
$H\to\gamma\gamma$ to $5\%$ for $H\to WW$ and $H\to\tau\tau$ to $10\%$
for $H\to b\bar{b}$ on the shape plus the statistical error of the
background sample used for normalization) and theoretical and
parametric uncertainties on Higgs boson production ($20\%$ GF, $15\%$
$t\bar{t}H$, $7\%$ $WH/ZH$, $4\%$ WBF) and decays ($1\%$, as a future
expectation).  For the WBF channels there is an additional uncertainty
on the minijet veto and jet tagging efficiency (combined) of $5\%$, as
well as an added $10\%$ contribution from $gg\to
Hgg$~\cite{DelDuca:2001eu}, which has its own theory uncertainty of a
factor of 2.

The $1\sigma$ uncertainties on each parameter are determined in the fit
by finding the maximum deviation of that parameter from its best
fit value that 
lies on the $\Delta \chi^2 = 1$ surface.  We repeat the procedure for
each Higgs mass value in the range $110\leq m_H\leq 190$ GeV in steps
of 10 GeV.

We perform the fits under three luminosity assumptions for the
LHC:\\[.3em]
      30 fb$^{-1}$ at each of two experiments, denoted 
\underline{$2 \times 30$~fb$^{-1}$};\\[.3em]
      300 fb$^{-1}$ at each of two experiments, of which only 100 fb$^{-1}$
is usable for WBF channels at each experiment, denoted 
\underline{$2 \times 300$ $+$ $2 \times 100$~fb$^{-1}$};\\[.3em]
      300 fb$^{-1}$ at each of two experiments, with the full luminosity
usable for WBF channels, denoted \underline{$2 \times
300$~fb$^{-1}$}.\\[.3em]
The second case allows for possible significant degradation of the WBF
channels in a high luminosity environment, while the third case serves
to investigate the possible physics gain of additional improvements in 
WBF studies at high luminosity.

In both cases the Higgs boson mass is not fitted, i.e.\ it is assumed
that the mass of the Higgs boson can be measured with high precision
($\Delta m_H/m_H<1\%$) in $H\to Z^{(*)}Z^{(*)}\to 4\ell$ or
$H\to\gamma\gamma$. If both channels go unobserved, the theoretical
predictions of Higgs boson branching ratios receive a large error due to the
relatively low precision and larger systematic errors of $m_H$
measurements in WBF $H\to\tau\tau$ or $H\to WW$. 


\section{Results for general multi-Higgs-doublet models}
\label{sec:results}

We obtain the results for the Higgs couplings-squared in general
multi-Higgs-doublet models using the assumption that
\begin{equation}
g^2(H,W)<1.05 \cdot g^2(H,W,SM), \quad g^2(H,Z)<1.05 \cdot g^2(H,Z,SM)~.
\end{equation}
Any model that contains only Higgs doublets and
singlets will satisfy the relations $g^2(H,W)\leq g^2(H,W,SM)$ and
$g^2(H,Z)\leq g^2(H,Z,SM)$.  The extra $5\%$ margin allows for
theoretical uncertainties in the translation between couplings-squared
and partial widths, and also for small admixtures of exotic Higgs
states, like SU(2) triplets.  As explained above, we allow for the 
possibility of
additional particles running in the loops for $H\to\gamma\gamma$ and
$gg\to H$, fitted by a positive or negative new partial width to these
contributions.

\begin{figure}[thb]
\begin{center}
\resizebox{\textwidth}{!}{
\includegraphics{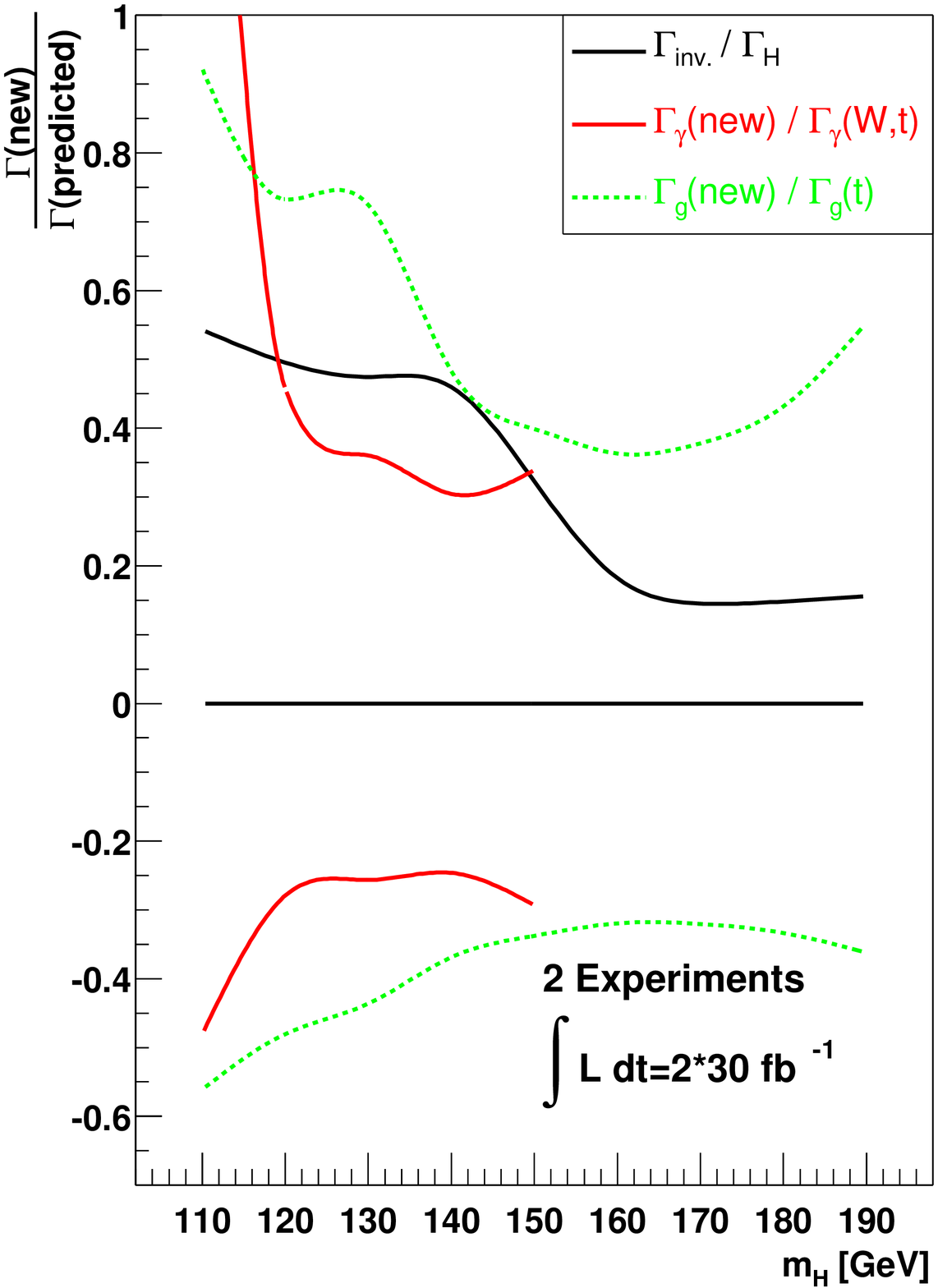}
\includegraphics{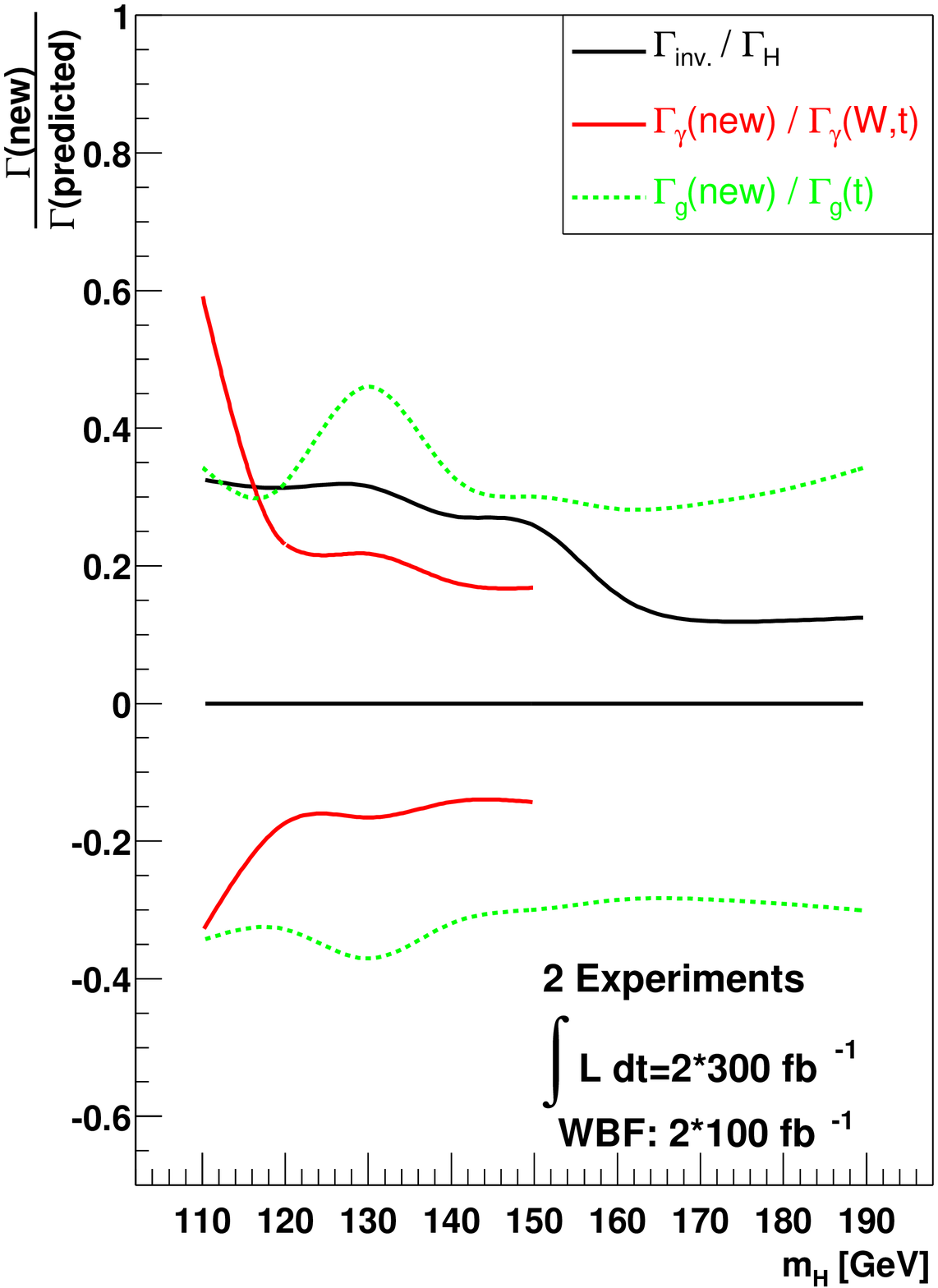}
}
\caption{Relative precisions of fitted new partial widths as a function
  of the Higgs mass assuming that SM rates are observed with 
  30~fb$^{-1}$ at each of two experiments
  (left) and 300~fb$^{-1}$ at each of two experiments for all channels
  except WBF, for which 100 fb$^{-1}$ is assumed (right).  The new
  partial width can be due to new particles in the loops for
  $H\to\gamma\gamma$ and $gg\to H$ or due to unobservable decay modes.
  See text for details.  Here we make the weak assumption that
  $g^2(H,V)<1.05 \cdot g^2(H,V,SM)$ ($V=W,Z$).}
\label{fig:Gnew}
\end{center}
\end{figure}

The results for the constraints on the new partial widths are shown in
Fig.~\ref{fig:Gnew} as a function of Higgs mass for the $2 \times 30$~fb$^{-1}$
and $2 \times 300 + 2 \times 100$~fb$^{-1}$ luminosity scenarios
assuming that SM rates are observed. The new 
partial
width for $H\to\gamma\gamma$ is most tightly constrained for $120\lsim
m_H\lsim 140$ GeV, being less than $\pm (25-35)\%$ of
$\Gamma_\gamma^{\rm SM}$ for $2 \times 30$~fb$^{-1}$ and $\pm (10-15)\%$ for
$2 \times 300 + 2 \times 100$~fb$^{-1}$. The new partial width for $gg\to H$ 
is less
well constrained, being less than $\pm (30-90)\%$ of $\Gamma_g^{\rm
  SM}$ for $2 \times 30$~fb$^{-1}$ and $\pm (30-45)\%$ for 
$2 \times 300 + 2 \times 100$~fb$^{-1}$ over the whole range of Higgs masses.

Additional light hadronic decays of the Higgs boson are fitted with a
partial width for undetected decays. (Invisible decays, e.g.\ to
neutralinos could still be observable~\cite{Eboli:2000ze}.)  This
undetected partial width can be constrained to be less than $15-55\%$
of the total fitted Higgs width for $2 \times 30$~fb$^{-1}$ and $15-30\%$ for
$2 \times 300 + 2 \times 100$~fb$^{-1}$, at the $1\sigma$ level.  
This undetected
partial width is most tightly constrained for Higgs masses above
160~GeV.

\begin{figure}[thb]
\begin{center}
\resizebox{\textwidth}{!}{
\includegraphics{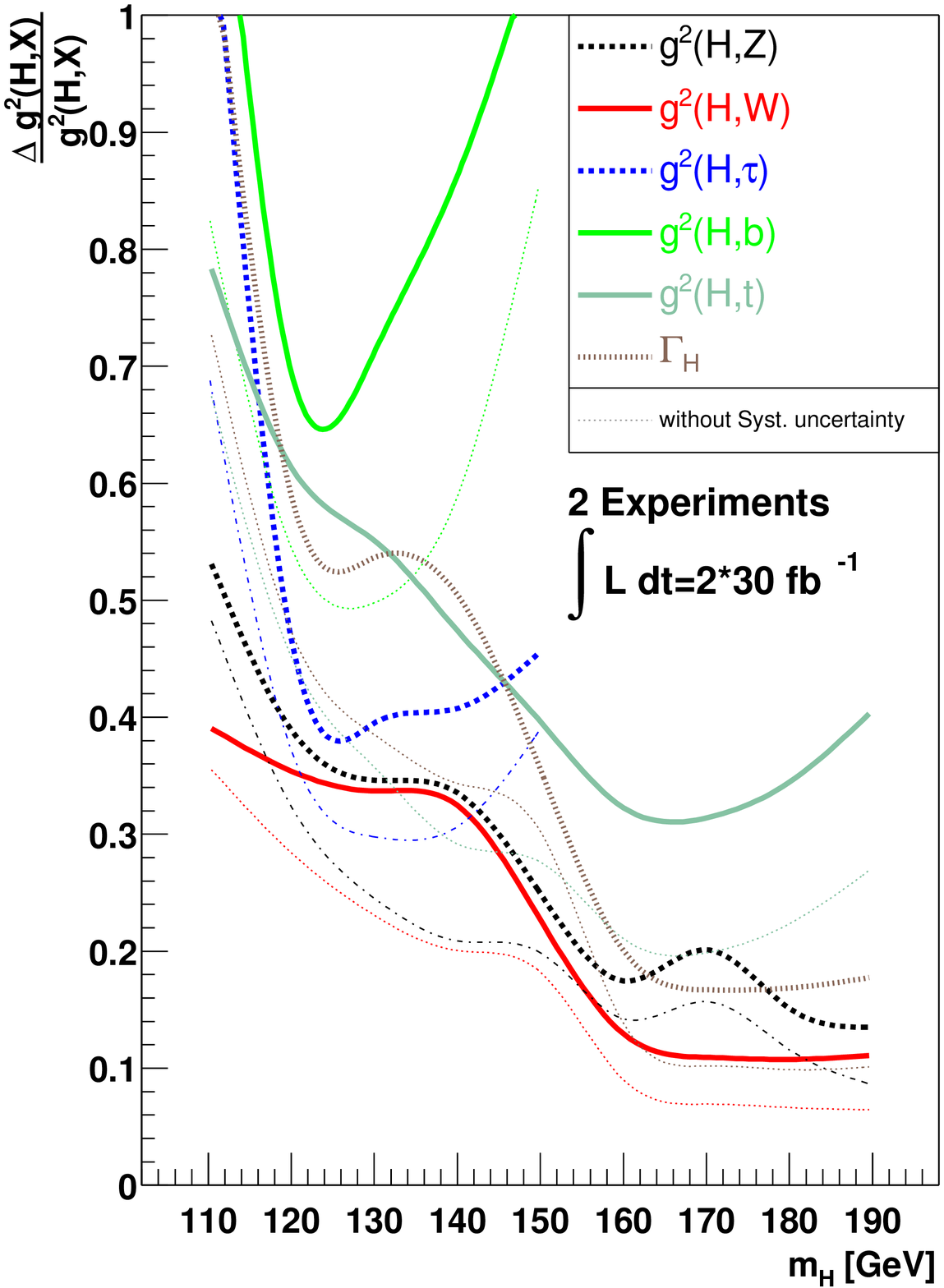}
\includegraphics{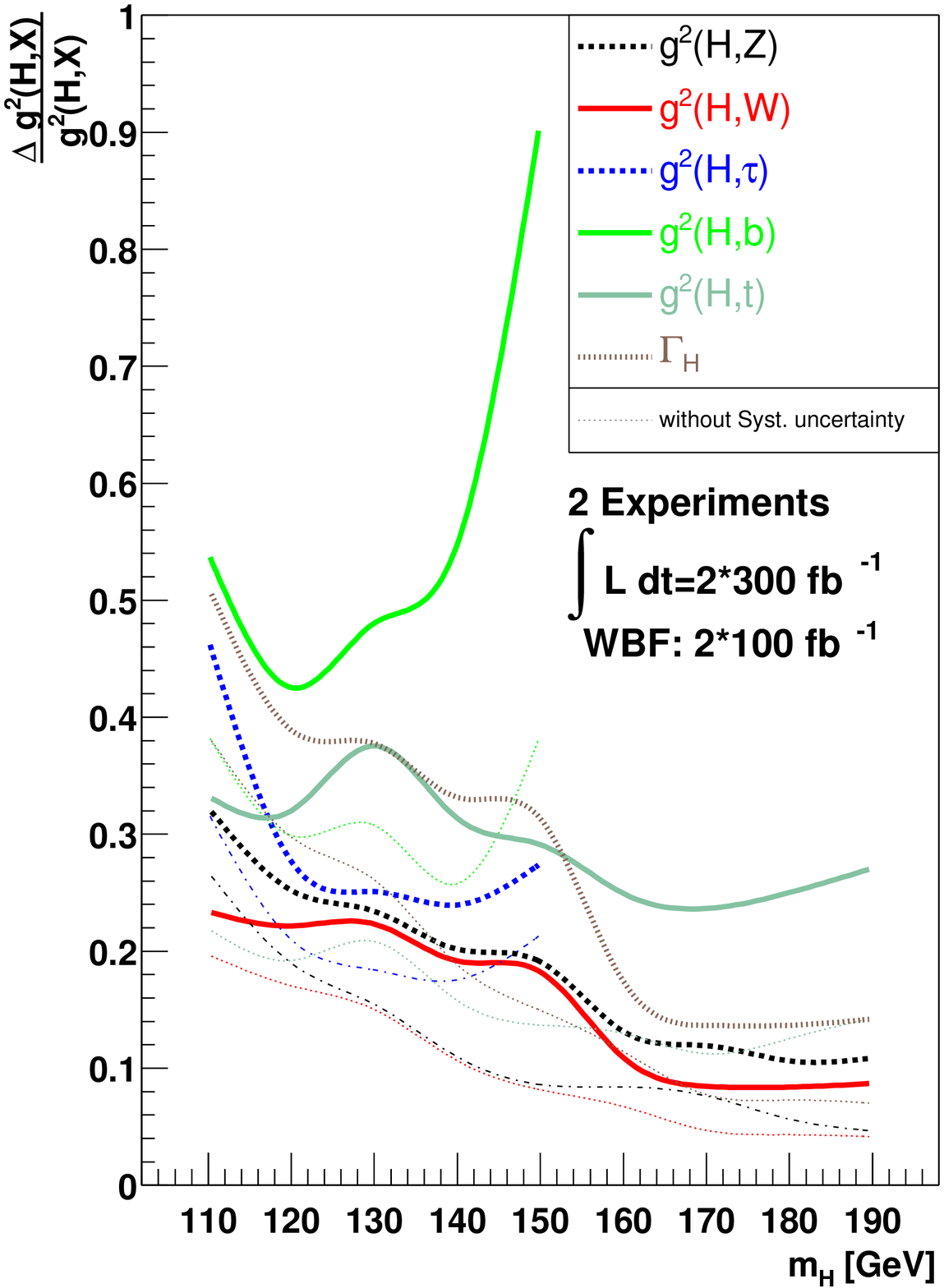}
}
\caption{Relative precision of fitted Higgs couplings-squared as a function
  of the Higgs mass for the $2 \times 30$~fb$^{-1}$ (left) and the 
  $2 \times 300 + 2 \times 100$~fb$^{-1}$ (right) luminosity scenarios.  
  We make the weak
  assumption that $g^2(H,V)<1.05 \cdot g^2(H,V,SM)$ ($V=W,Z$) but
  allow for new particles in the loops for $H\to\gamma\gamma$ and
  $gg\to H$ and for unobservable decay modes. See text for details.}
\label{fig:fit}
\end{center}
\end{figure}

The resulting precisions on the Higgs boson couplings squared are
shown in Fig.~\ref{fig:fit} as a function of Higgs mass for the same
luminosity scenarios, $2 \times 30$~fb$^{-1}$ and 
$2 \times 300 + 2 \times 100$~fb$^{-1}$, assuming SM rates.  For
$2 \times 300 + 2 \times 100$~fb$^{-1}$,
typical accuracies range between 20 and $30\%$ for
Higgs masses below 150~GeV.  Above $W$-pair threshold the measurement
of the then-dominant $H\to WW,ZZ$ partial widths improves to the
$10\%$ level.  The case of $2 \times 300$~fb$^{-1}$ yields only small
improvements over the right-hand panel in Fig.~\ref{fig:fit}, except
in the case of $g^2(H,\tau)$ which shows moderate improvement.
However, since this happens for Higgs masses below $\sim 140$~GeV,
this effect can be relatively important in the case of MSSM analyses,
see Sec.~\ref{sec:mssm_specific}.  This can be understood because the
$H\to\tau\tau$ decay is measured only in WBF, and $g(H,\tau)$ does not
have a large effect on the Higgs total width or loop-induced
couplings.

The results shown in Fig.~\ref{fig:fit} reflect present understanding
of detector effects and systematic errors. One should note that
improved selection and higher acceptance will decrease the statistical
errors.  At least as important is work on the reduction of systematic
errors. In Fig.~\ref{fig:fit}, the thin lines show expectations with
vanishingly small systematics: systematic errors contribute up to half
the total error, especially at high luminosity.

For a Higgs boson mass below 140 GeV the main contribution to the
systematic uncertainty is the background normalization from sidebands.
The largest contribution is from $H\to b\bar{b}$. For this channel the
signal to background ratio is between 1:4 and 1:10. For the background
normalization we assume a systematic error of
$10\%$~\cite{ATL-PHYS-2003-024}. This leads to a huge total systematic
error on the measurement of $\Gamma_b$, which is the main contribution
to the total width $\Gamma_H$ (the BR($H \to b\bar{b}$) is between
$80\%$ and $30\%$). But a measurement of absolute couplings needs
$\Gamma_H$ as input, as discussed above,
so all measurements of couplings share the large systematic
uncertainty on $H \to b\bar{b}$.

For a Higgs boson mass above 150~GeV there are two dominant
contributions to the systematic error:
the background normalizations in GF, WBF and $t\bar{t}H$
  (systematic error between $5\%$ and $15\%$)
and the QCD uncertainty in the cross section calculations for GF
  ($20\%$) and $t\bar{t}H$ ($15\%$) from given Higgs boson couplings.
This is especially evident in the measurement of the top coupling
which is based on the $t\bar{t}H$ channel. Here the systematic uncertainties
contribute half of the total error.

The precision of the extracted couplings improves if more restrictive
theoretical assumptions are applied, see 
Refs.~\cite{Hcoupl,Assamagan:2004mu} for a discussion.


\section{Sensitivity to deviations from the Standard Model}
\label{sec:mssm_specific}

If the values obtained for the Higgs boson couplings differ from the SM
predictions, one can investigate at which significance the SM can be
excluded from LHC measurements in the Higgs sector alone. As a specific
example of physics beyond the SM, we consider here the MSSM.

If supersymmetric partners of the SM particles were detected at the LHC,
this would of course rule out the SM. It would nevertheless be of
interest in such a situation to directly verify the non-SM nature of the
Higgs sector. Besides the possible detection of the additional states
of an extended Higgs sector, a precise measurement of the couplings of
the lightest (SM-like) Higgs boson will be crucial. 

For definiteness let us assume that the pseudoscalar Higgs and
the charged Higgs are fairly heavy ($M_A\gsim 150$~GeV, and they may,
but need not, have been observed directly) so that they do not
interfere with the $h$~signal extraction. We furthermore assume that
only decays into SM particles are detected. A fit of the Higgs couplings
can then be performed as outlined above, where the rates are obtained 
according to a certain MSSM scenario.
A quantitative, global measure of how well the LHC can distinguish the
SM from a specific MSSM scenario is provided by a $\chi^2$-analysis of
the deviations expected for this SUSY scenario. 

As a specific
example we consider the no-mixing scenario scenario of
Ref.~\cite{Carena:2002qg} (for results in the other benchmark scenarios
of Ref.~\cite{Carena:2002qg}, see Ref.~\cite{Hcoupl}). We calculate the 
mass and branching
fractions of the MSSM Higgs boson using HDECAY3.0
\cite{Djouadi:1998yw}, using the FeynHiggsFast1.2.2
\cite{Heinemeyer:2000nz,Heinemeyer:1999be} option to compute the MSSM
Higgs masses and couplings. Assuming that, for a given $M_A$ and
$\tan\beta$, the corresponding SUSY model is realized in nature, we
may ask at what significance the SM would be ruled out from
$h$~measurements alone. 

\begin{figure}[htb!]
\begin{center}
\resizebox{\textwidth}{!}{
\rotatebox{270}{\includegraphics[50,50][555,590]{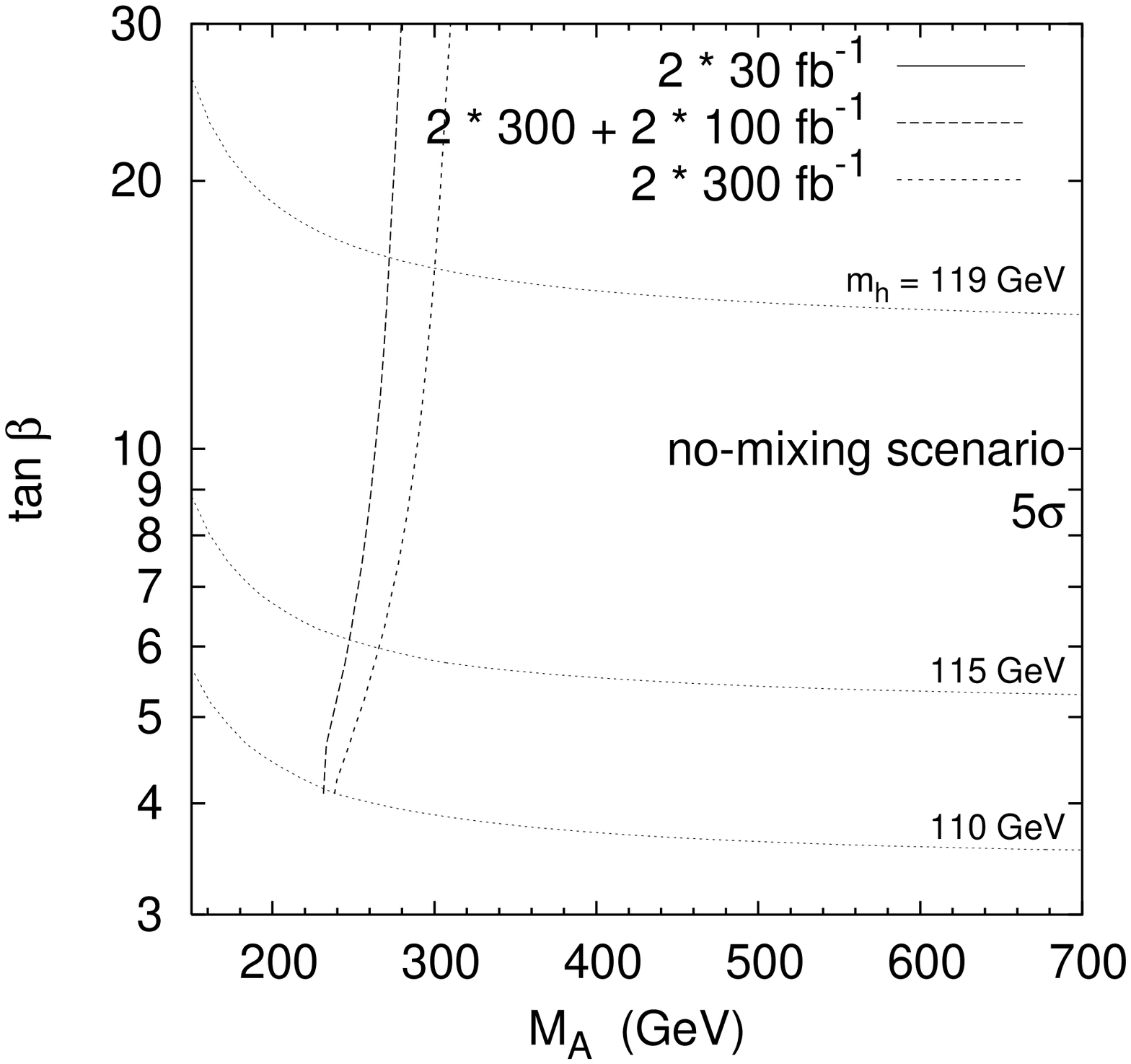}}
\rotatebox{270}{\includegraphics[50,50][555,590]{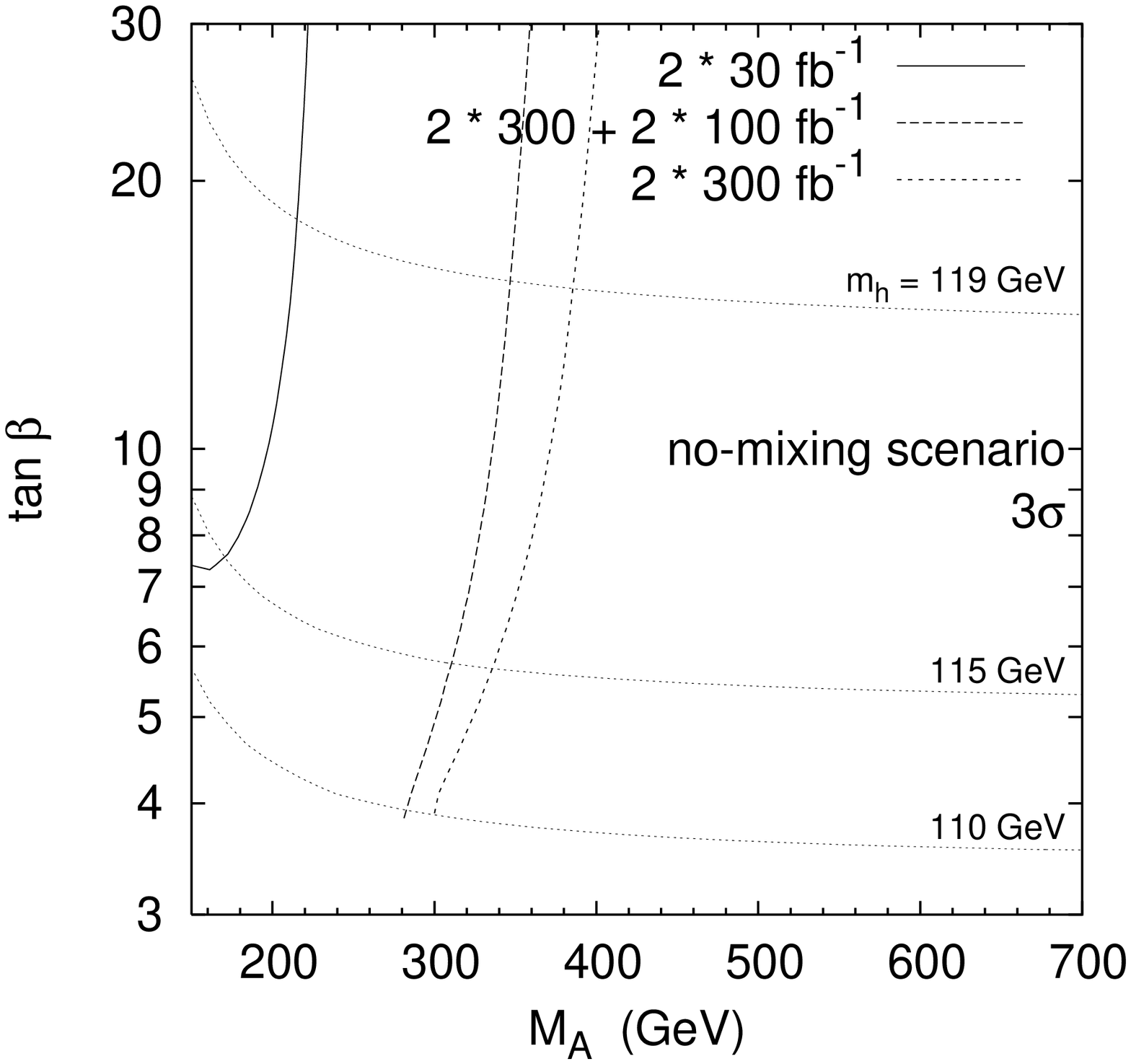}}
}
\caption{Fit within the no-mixing benchmark scenario of the MSSM 
  in the 
  $M_A$--$\tan\beta$ plane for three luminosity scenarios.  The two
  panels show the region (to the left of the curves) which would yield
  a $\geq 5\sigma$ ($\Delta\chi^2\geq 25$) or $\geq 3\sigma$
  ($\Delta\chi^2\geq 9$) discrepancy from the SM.  The
  mostly-horizontal dotted lines are contours for different values 
  of~$m_h$. }
\label{fig:mssm_contour_nomix}
\end{center}
\end{figure}

The resulting contours are shown in
Fig.~\ref{fig:mssm_contour_nomix} for the three luminosity assumptions
defined above.  In the areas to the left of the
contours the SM can be rejected with more than $5\sigma$ or $3\sigma$
significance, respectively.
The $\chi^2$ definition in Fig.~\ref{fig:mssm_contour_nomix} assumes the
same systematic errors as our analysis in Sec.~\ref{sec:results}.
Event rates and resulting statistical errors, however, are those
expected for the MSSM. 
For $2 \times 300 + 2 \times 100$~fb$^{-1}$ a deviation from the SM can
be established at the $3\sigma$ level in this scenario up to 
$M_A\simeq 350$~GeV and at the $5\sigma$ level up to $M_A\simeq 250$~GeV.

The source of the MSSM analysis sensitivity can be understood as
follows.  For $M_A\gsim 200$~GeV, the
couplings of $h$ to SM particles all essentially obtain their SM
values except for the $hbb$ and $h\tau\tau$ couplings, due to the
slower decoupling behavior of the latter.  In this scenario the SUSY
threshold corrections to the $b$ mass are also quite small, so that
the ratio of the $hbb$ and $h\tau\tau$ couplings essentially takes its
SM value.  The $h\to b\bar{b}$ decay mode dominates the Higgs total
width in this scenario.  
The pattern of Higgs coupling deviations can
then be summarized as follows: all the Higgs production cross sections
considered in our study are SM-like. The partial widths into
$b\bar{b}$ and $\tau\tau$ are equally enhanced (but with SM-like
banching ratios since the total width is dominated by
$b\bar{b}$ and $\tau\tau$ decays).
This results in a larger total width for the Higgs boson. The
branching ratios into all other final states ($WW^*$, $ZZ^*$,
$\gamma\gamma$) are smaller than in the SM, reflecting this
total width enhancement.


It should be noted that the shown sensitivity to $M_A$ cannot directly
be translated into indirect bounds on $M_A$. In order to establish
realistic bounds on $M_A$, a careful analysis of the experimental errors 
arising from the incomplete knowledge of the spectrum of supersymmetric 
particles and of the theoretical uncertainties from unknown higher-order
corrections is necessary.


\section{Conclusions}
\label{sec:sum_out}

Measurements of the Higgs sector are expected to provide many
complementary signatures after several years of LHC running. Combining
these measurements allows one to extract information on Higgs partial
widths and Higgs couplings to fermions and gauge bosons. Because
significant contributions from unobservable channels cannot easily be
ruled out at the LHC, model-independent analyses produce large
correlations between extracted partial widths. We have shown that a 
reduction of correlations and hence an absolute determination of Higgs
boson couplings and the total Higgs width can be
achieved if weak theory assumptions are made.  We have
analyzed the constraint valid in generic multi-Higgs-doublet
models, namely that $HVV$ couplings cannot be larger than within the
SM. Within such models, the LHC can measure Higgs couplings to the top
quark, tau lepton, and $W$ and $Z$ bosons with accuracies in the
$10-30\%$ range once 300~fb$^{-1}$ of data have been collected.  If,
on the other hand, the SLHC will be realized, one could hope for
significant improvements over the results presented here. This applies 
in particular for the bottom Yukawa coupling determination.

Within the MSSM, significant deviations in the Higgs sector should be
observable at the LHC, provided that the charged and the pseudoscalar
Higgs masses are not too heavy, i.e., that decoupling is not
completely realized. Within the no-mixing benchmark scenario and with
300~fb$^{-1}$ of data, the LHC can distinguish the MSSM and the SM at
the $3\sigma$ level up to $M_A\simeq 350$~GeV and with $5\sigma$
significance up to $M_A\simeq 250$~GeV with the Higgs data alone.  The
LHC will thus provide a surprisingly sensitive first look at the Higgs
sector, even though it cannot match the precision and
model-independence of analyses which are expected for a linear
$e^+e^-$
collider~\cite{Aguilar-Saavedra:2001rg,Abe:2001np,Abe:2001gc}.

So far we have investigated the situation where no important channel
suffers substantial suppression. Scenarios where such a
suppression occurs will be analyzed in a forthcoming publication.

 
\section*{Acknowledgments}
We would like to thank M.~Carena for enjoyable discussions during
early stages of this work.  H.L.\ was supported in part by the
U.S.~Department of Energy under grant DE-FG02-95ER40896 and in part by
the Wisconsin Alumni Research Foundation.  
This work has been supported by the European Community's Human
Potential Programme under contract HPRN-CT-2000-00149 Physics at
Colliders.
G.W.\ thanks the organizers of the
XXXIXth Rencontres de Moriond for the kind invitation and the pleasant
atmosphere enjoyed at La Thuile.


\end{document}